\documentclass[nofootinbib,preprint,preprintnumbers,amsmath,amssymb]{revtex4}

\usepackage{graphicx}

\newcommand{\GeV}{\,{\rm GeV}}
\newcommand{\MeV}{\,{\rm MeV}}
\newcommand{\TeV}{\,{\rm TeV}}
\newcommand{\s}{\,{\rm s}}
\newcommand{\sr}{\,{\rm sr}}
\newcommand{\m}{\,{\rm m}}

\begin{document}

\preprint{DESY 09-083}
\preprint{TUM-HEP 724/09}

\title{Decaying Dark Matter in Light \\of the PAMELA and Fermi LAT Data}

\author{Alejandro Ibarra}
\email{alejandro.ibarra@ph.tum.de}
\author{David Tran}
\email{david.tran@ph.tum.de}
\affiliation{
Physik-Department T30d, Technische Universit\"at M\"unchen,\\
James-Franck-Stra\ss e, 85748 Garching, Germany.
}

\author{Christoph Weniger}
\email{christoph.weniger@desy.de}
\affiliation{
Deutsches Elektronen-Synchrotron DESY,\\
Notkestra\ss e 85, 22607 Hamburg, Germany.
}

\begin{abstract}
A series of experiments measuring high-energy cosmic rays have recently
reported strong indications for the existence of an excess of high-energy
electrons and positrons. If interpreted in terms of the decay of dark matter
particles, the PAMELA measurements of the positron fraction and the Fermi LAT
measurements of the total electron-plus-positron flux restrict the possible
decaying dark matter scenarios to a few cases.  Analyzing different decay
channels in a model-independent manner, and adopting a conventional diffusive
reacceleration model for the background fluxes of electrons and positrons, we
identify some promising scenarios of dark matter decay and calculate the
predictions for the diffuse extragalactic gamma-ray flux, including the
contributions from inverse Compton scattering with the interstellar radiation
field.
\end{abstract}


\maketitle

\section{Introduction}
Different experiments measuring high-energy cosmic rays have over the last
months reported a wealth of new results pointing to the existence of an exotic
source of electrons and positrons. The PAMELA collaboration reported evidence
for a sharp rise of the positron fraction at energies $7-100$
GeV~\cite{Adr08}, possibly extending toward even higher energies, compared to
the expectations from spallation of primary cosmic rays on the interstellar
medium~\cite{MS98}. This result confirmed previous hints about the existence
of a positron excess from HEAT~\cite{Bar97}, CAPRICE~\cite{Boe00} and
AMS-01~\cite{Agu07}. Almost at the same time, the balloon-borne experiments
ATIC~\cite{ATIC} and PPB-BETS~\cite{Tor08} reported the discovery of a peak in
the total electron-plus-positron flux at energies $600-700$ GeV, while the
H.E.S.S. collaboration~\cite{Aha08} reported a substantial steepening in the
high-energy electron-plus-positron spectrum above 600 GeV compared to lower
energies. 

These results raised a lot of interest in the astrophysics and particle
physics communities, leading to many proposals trying to explain this excess.
One of the most popular astrophysical interpretations of the positron excess
is in terms of the electron-positron pairs produced by the interactions of
high-energy photons in the strong magnetic field of
pulsars~\cite{pulsars,HBS08,Gri04}. However, this interpretation requires a
rather large fraction of the spin-down power being injected in the form of
electron-positron pairs or a rather large rate of gamma-ray pulsar formation.
Alternatively, the positrons could be originating from the decay of charged
pions, which are in turn produced by the hadronic interactions of high-energy
protons accelerated by nearby sources~\cite{Bla09}. 

An arguably more exciting explanation of the cosmic-ray positron excess is the
possibility that the positrons are produced in the annihilation or the decay
of dark matter particles. Should this interpretation be confirmed by future
experiments, then the positron excess would constitute the first
non-gravitational evidence for the existence of dark matter in our Galaxy. The
interpretation of the PAMELA excess in terms of dark matter is subject to
constraints from the flux measurements of other cosmic-ray species. A very
important constraint arises from the measurements of the antiproton flux by
PAMELA~\cite{Adr08b}, BESS95~\cite{Mat98}, BESS95/97~\cite{Ori99},
CAPRICE94~\cite{Boe97}, CAPRICE98~\cite{Boe01} and IMAX~\cite{Mit96}, which
are consistent with the expectations from conventional propagation models,
thus excluding the possibility of a large antiproton flux from dark matter
annihilation or decay~\cite{Ber99,Don01}. 

The steep rise in the positron fraction observed by PAMELA can be explained by
dark matter annihilations in the center of the Milky Way, provided the dark
matter particle has a mass larger than $\sim 150$ GeV and annihilates
preferentially into leptons of the first or second generation~\cite{CKR09}.
This interpretation of the positron excess, however, typically requires the
{\it ad hoc} introduction of large boost factors. Furthermore, it has been
argued that if dark matter annihilations are the origin of the PAMELA anomaly,
then the predicted gamma-ray emission from the center of the Galaxy is in
conflict with the H.E.S.S. observations for typical cuspy halo
profiles~\cite{BCS09}. On the other hand, if the positron excess is due to the
decay of dark matter particles, the dark matter particles must have a mass
larger than $\sim 300$ GeV, a lifetime around $10^{26}$ s, and must decay
preferentially into hard leptons of the first or second
generation~\cite{IT09}. In this case, no boost factors are required and the
gamma and radio measurements are consistent with present
measurements~\cite{NST09}. Some recent works on the indirect detection of
decaying dark matter can be found in~\cite{IT08,BBCI,decay,gaugino}.
 
More recently, the Fermi LAT collaboration has published measurements of the
electron-plus-positron flux from 20 GeV to 1 TeV of unprecedented
accuracy~\cite{Abd09}, revealing an energy spectrum that roughly follows a
power law $\propto E^{-3.0}$ without any prominent spectral features.
Simultaneously, the H.E.S.S. collaboration reported a measurement of the
cosmic-ray electron-plus-positron spectrum at energies larger than 340 GeV,
confirming the Fermi result of a power-law spectrum with spectral index of
$3.0 \pm 0.1 {\rm (stat.)} \pm 0.3 {\rm (syst.)}$, which furthermore steepens
at about 1 TeV ~\cite{Aha09}.  The measured energy spectrum is harder than
expected from conventional diffusive models, although it can be accommodated
by an appropriate change of the injection spectrum of primary electrons.
However, when taken together with the steep rise in the positron fraction as
seen by PAMELA up to energies of 100 GeV, the Fermi LAT data suggest the
existence of additional Galactic sources of high-energy electrons and
positrons with energies up to a few TeV. Furthermore, it should be borne in
mind that the determination of the correct Galactic cosmic-ray scenario is
still an open problem, and while an electron injection spectrum harder than
the conventional could reproduce the Fermi data, it fails to account for the
AMS-01 and HEAT data below 20 GeV and the H.E.S.S. data above 1
TeV~\cite{Gra09}.

In this paper we analyze the constraints that the results of the PAMELA and
Fermi collaborations impose on the scenario of decaying dark matter, assuming
a GALPROP conventional model as our Galactic cosmic-ray scenario.  To this
end, we pursue a model-independent approach, calculating the prediction for
the positron fraction and the total electron-plus-positron flux for various
decay channels of both a fermionic and a bosonic dark matter particle. We will
identify the most promising scenarios in the light of the PAMELA and Fermi
data, and we will calculate for those the predictions for the antiproton flux
and the diffuse extragalactic gamma-ray flux. Some related works have recently
appeared~\cite{Meade, related}.

The paper is organized as follows: in Section 2 we will review the production
and propagation in the Galaxy of high-energy electrons/positrons, antiprotons
and gamma rays from dark matter decay, including a contribution to the total
gamma-ray flux from inverse Compton radiation. In Section 3 we will show the
predictions for the positron fraction and the total electron-plus-positron
flux for several decaying dark matter scenarios. For the promising scenarios,
we will also show the predictions for the antiproton and the gamma-ray fluxes.
Finally, in Section 4 we will present our conclusions.

\section{Cosmic rays} 
In this section, we briefly review the propagation model for cosmic rays that
we need for the calculation of the electron, positron and antiproton fluxes
measurable at Earth. Furthermore, we discuss our calculation of the flux of
gamma rays, which come from inverse Compton scattering (ICS) with the
interstellar radiation field (ISRF) as well as directly from the decay process
itself.

If dark matter decays at a sufficiently large rate, the decay products
(electrons, positrons, antiprotons and gamma rays) could be observable as an
anomalous contribution to the high-energy cosmic-ray fluxes. The production
rate of particles per unit energy and unit volume at a position $\vec{r}$ with
respect to the center of the Milky Way is given by
\begin{equation}
  Q(E,\vec{r})=\frac{\rho(\vec{r})}{M_{\rm DM}\,\tau_{\rm DM}}\frac{dN}{dE}\;,
  \label{source-term}
\end{equation}
where $dN/dE$ is the energy spectrum of particles produced in the decay and
$\rho(\vec{r})$ is the density profile of dark matter particles in the Milky
Way halo. For definiteness we will adopt the spherically symmetric
Navarro-Frenk-White halo density profile~\cite{NFW96}:
\begin{equation}
  \rho(r)=\frac{\rho_0}{(r/r_c)
  [1+(r/r_c)]^2}\;,
\end{equation}
with $\rho_0\simeq 0.26\,{\rm GeV}/{\rm cm}^3$ and $r_c\simeq 20 ~\rm{kpc}$,
although our results are almost independent of choice of the density
profile\footnote{Due to the effective energy loss of electrons, the
high-energy component of the spectrum mostly originates from sources within
the Galactic neighborhood of a few kpc from the Solar System, where the
different halo profiles are very similar. We have checked that choosing
different halo profiles has a negligible effect on our results (see also
\cite{IT08}).}.

\subsection{Electron/positron propagation}
After being produced in the Milky Way halo, the electrons and positrons
propagate through the Galaxy and its diffusive halo in a rather complicated
way before reaching the Earth. The propagation is commonly described by a
stationary two-zone diffusion model with cylindrical boundary
conditions~\cite{ACR}. Under this approximation, the number density of
electrons and positrons per unit energy, $f_{e^\pm}(E,\vec{r},t)$, satisfies
the following transport equation:
\begin{equation}
  0=\frac{\partial f_{e^\pm}}{\partial t}=
  \nabla \cdot [K(E,\vec{r})\nabla f_{e^\pm}] +
  \frac{\partial}{\partial E} [b(E,\vec{r}) f_{e^\pm}]+Q_{e^\pm}(E,\vec{r})\;.
  \label{transport1}
\end{equation}
The first term on the right-hand side of the transport equation is the
diffusion term, which accounts for the propagation through the tangled
Galactic magnetic field. The diffusion coefficient $K(E,\vec{r})$ is assumed
to be constant throughout the diffusion zone and is parametrized by $K(E)=K_0
\;\beta\; {\cal R}^\delta$, where $\beta=v/c$ and ${\cal R}$ is the rigidity
of the particle, which is defined as the momentum in GeV per unit charge,
${\cal R}\equiv p({\rm GeV})/Z$. The second term accounts for energy losses
due to ICS on starlight or the cosmic microwave background (CMB), synchrotron
radiation and ionization. We parameterize the energy loss rate as $b(E) =
\frac{E^2}{E_0 \tau_E}$, with $E_0 = 1\GeV$ and $\tau_E = 10^{26} \s$.
Lastly, $Q_{e^\pm}(E,\vec{r})$ is the source term of electrons and positrons,
defined in Eq.~(\ref{source-term}).

The boundary conditions for the transport equation, Eq.(\ref{transport1}),
require the solution $f_{e^\pm}(E,\vec{r},t)$ to vanish at the boundary of the
diffusion zone, which is approximated by a cylinder with half-height $L =
1-15~\rm{kpc}$ and radius $ R = 20 ~\rm{kpc}$. Under these assumptions, the
propagation of electrons and positrons can be described by just three
parameters, the normalization $K_0$ and the spectral index $\delta$ of the
diffusion coefficient, which are related to the properties of the interstellar
medium, and the height of the diffusion zone, $L$. In our numerical analysis
we will adopt for these parameters the values of the MED propagation model
defined in \cite{MDT+01}, which provide the best fit to the Boron-to-Carbon
(B/C) ratio: $\delta=0.70$, $K_0=0.0112\,{\rm kpc}^2/{\rm Myr}$ and $L=4\,{\rm
kpc}$. Our conclusions, however, are rather insensitive to the choice of
propagation parameters, as the different sets of propagation parameters yield
rather similar results for cosmic rays from local sources. This is due to the
fact that at higher energies above several $10\GeV$ energy losses dominate the
effects of diffusion, rendering the exact propagation model parameters less
relevant. We illustrate the dependence of the results on the adopted model
parameters for a particular example in Section 3.

The solution of the transport equation at the heliospheric boundary,
$r=r_\odot$, $z=0$, can be formally expressed by the convolution
\begin{equation}
  f_{e^\pm}(E)=\frac{1}{M_{\rm DM}\, \tau_{\rm DM}}
  \int_0^{M_{\rm DM}}dE^\prime\; G_{e^\pm}(E,E^\prime) \,
  \frac{dN_{e^\pm}(E^\prime)}{dE^\prime}\;.
  \label{solution}
\end{equation}
The Green's function $G_{e^\pm}(E,E^\prime)$ encodes all the information about
astrophysics (such as the details of the halo profile and the propagation of
electrons/positrons in the Galaxy), while the remaining part is
model-dependent and is determined by the nature of the dark matter particles.
Analytical and numerical expressions for the Green's function for the
propagation of electrons/positrons can be found in~\cite{IT08}.

Finally, the interstellar flux of primary electrons/positrons from dark matter
decay is given by:
\begin{equation}
  \Phi_{e^\pm}^{\rm{DM}}(E) = \frac{c}{4 \pi} f_{e^\pm}(E)\;.
  \label{flux-positron}
\end{equation}

In order to compare our results with the PAMELA results of the positron
fraction as well as the Fermi results on the total flux of electrons plus
positrons it is necessary to know the background fluxes of high-energy
electrons and positrons. The background flux of positrons is constituted by
secondary positrons produced in the collision of primary protons and other
nuclei with the interstellar medium. On the other hand, the background flux of
electrons is constituted by a primary component, presumably produced in
supernova remnants, as well as a secondary component, produced by spallation
of cosmic rays on the interstellar medium and which is much smaller than the
primary component. Whereas the spectrum of secondary electrons and positrons
is calculable in a given propagation model, the energy spectrum and the
normalization of the primary electron flux is unknown and has to be determined
by direct measurements. 

In this paper we will adopt for the background fluxes of electrons and
positrons the ones corresponding to the ``model 0'' presented by the Fermi
collaboration in \cite{Gra09}, which fits well the low-energy data points of
the total electron-plus-positron flux and the positron fraction.  The
interstellar background fluxes can be parametrized as:
\begin{eqnarray}
  \Phi^{\rm bkg}_{e^-}(E)&=&
  \left(\frac{82.0\,\epsilon^{-0.28}}{1+0.224\,\epsilon^{2.93}}\right)
  \GeV^{-1}\m^{-2}\s^{-1}\sr^{-1}\;,\\
  \Phi^{\rm bkg}_{e^+}(E)&=& \left(\frac{38.4
  \,\epsilon^{-4.78}}{1+0.0002\,\epsilon^{5.63}}
  +24.0\,\epsilon^{-3.41}\right)
  \GeV^{-1}\m^{-2}\s^{-1}\sr^{-1}\;,
  \label{eqn:Backgrounds}
\end{eqnarray}
where $\epsilon=E/1\GeV$. In the energy regime between $2\GeV$ and $1\TeV$
these approximations are better than $5\%$.

At energies smaller than $\sim 10$ GeV the electron/positron fluxes at the top
of the atmosphere can differ considerably from the interstellar fluxes, due to
solar modulation effects. Under the force field
approximation~\cite{solar-modulation}, the fluxes at the top of the atmosphere
are related to the interstellar fluxes by the following simple
relation~\cite{perko}:
\begin{equation}
  \Phi_{e^\pm}^{\rm TOA}(E_{\rm TOA})=
  \frac{E_{\rm TOA}^2}{E_{\rm IS}^2}\;
  \Phi_{e^\pm}^{\rm IS}(E_{\rm IS}),
\end{equation}
where $E_{\rm IS}=E_{\rm TOA}+\phi_F$, with $E_{\rm IS}$ and $E_{\rm TOA}$
being the electron/positron energies at the heliospheric boundary and at the
top of the Earth's atmosphere, respectively, and $\phi_F$ being the solar
modulation parameter, which varies between 500 MV and 1.3 GV over the
eleven-year solar cycle. In order to compare our predictions with the AMS-01
and HEAT data we will take $\phi_F=550$ MV~\cite{Bar97}.

Then, if there exists an exotic source of electrons and positrons from dark
matter decay, the total flux of electrons plus positrons and the positron
fraction at the top of the atmosphere read, respectively,
\begin{eqnarray}
  \Phi^{\rm tot}(E)&=&\Phi_{e^-}^{\rm{DM}}(E) +\Phi_{e^+}^{\rm{DM}}(E)
  + k \;  \Phi_{e^-}^{\rm{bkg}}(E) + 
  \Phi_{e^+}^{\rm{bkg}}(E)\;,\\
  {\rm PF}(E) &= &\frac{\Phi_{e^+}^{\rm{DM}}(E) + \Phi_{e^+}^{\rm{bkg}}(E)}
  {\Phi^{\rm tot}(E)},
\end{eqnarray}
where we have left the normalization of the primary electron flux as a free
parameter, $k$, to be determined in order to provide a qualitatively good fit
to the PAMELA and Fermi measurements. 

\subsection{Antiproton propagation} 
Antiproton propagation in the Galaxy can be described in a similar manner as
that of electrons and positrons. However, since antiprotons are much heavier
than electrons and positrons, energy losses are negligible. However,
antiproton propagation is affected by convection, which accounts for the drift
of antiprotons away from the disk induced by the Milky Way's Galactic wind.
Following \cite{MDT+01} we will assume that it has axial direction and that it
is constant inside the diffusion region: $\vec{V}_c(\vec{r})=V_c\; {\rm
sign}(z)\; \vec{k}$. Then, the transport equation for antiprotons reads:
\begin{equation}
  0=\frac{\partial f_{\bar p}}{\partial t}=
  \nabla \cdot [K(T,\vec{r})\nabla f_{\bar p}-
  \vec{V_c}(\vec{r})  f_{\bar p}] +Q_{\bar p}(T,\vec{r})\;,
  \label{transport2}
\end{equation}
where $T$ is the antiproton kinetic energy.

As for the case of electrons and positrons, the solution of the transport
equation at the heliospheric boundary, $r=r_\odot$, $z=0$, can be formally
expressed by the convolution
\begin{equation}
  f_{\bar p}(T)=\frac{1}{M_{\rm DM}\; \tau_{\rm DM}}
  \int_0^{T_{\rm max}}dT^\prime\; G_{\bar p}(T,T^\prime) \;
  \frac{dN_{\bar p}(T^\prime)}{dT^\prime}\;,
  \label{solution2}
\end{equation}
where $T_{\rm max}=M_{\rm DM}-m_p$ and $m_p$ is the proton mass. Analytical
and numerical expressions for the Green's function $G_{\bar p}(T,T^\prime)$
can be found in~\cite{IT08}. Finally, the interstellar flux of primary
antiprotons from dark matter decay is given by
\begin{equation}
  \Phi_{\bar p}^{\rm{DM}}(T) = \frac{v}{4 \pi} f_{\bar p}(T)\;,
  \label{flux-antiproton}
\end{equation}
where $v$ is the velocity of the antiprotons. The prediction of the antiproton
flux from dark matter decay is very sensitive to the choice of propagation
parameters. Therefore, we will show the results for three different
propagation models that are consistent with the observed B/C ratio and that
give the maximal (MAX), median (MED) and minimal (MIN) antiproton
flux~\cite{MDT+01}. The relevant parameters are summarized in
Tab.~\ref{tab:param-antiproton}.

\begin{table}[t]
  \begin{center}
    \begin{tabular}{|c|cccc|}
      \hline
      Model & $\delta$ & $K_0\,({\rm kpc}^2/{\rm Myr})$ & $L\,({\rm kpc})$
      & $V_c\,({\rm km}/{\rm s})$ \\
      \hline 
      MIN & 0.85 & 0.0016 & 1 & 13.5 \\
      MED & 0.70 & 0.0112 & 4 & 12 \\
      MAX & 0.46 & 0.0765 & 15 & 5 \\
      \hline
    \end{tabular}
    \caption{\label{tab:param-antiproton} \small Astrophysical parameters
    compatible with the B/C ratio that yield the minimal (MIN), median (MED)
    and maximal (MAX) flux of antiprotons.}
  \end{center}
\end{table}

The antiproton flux at Earth is also affected at low energies by solar
modulation effects. Again, under the force field
approximation~\cite{solar-modulation}, the antiproton flux at the top of the
Earth's atmosphere is related to the interstellar antiproton flux~\cite{perko}
by the simple relation:
\begin{equation}
  \Phi_{\bar p}^{\rm TOA}(T_{\rm TOA})=
  \left(
  \frac{2 m_p T_{\rm TOA}+T_{\rm TOA}^2}{2 m_p T_{\rm IS}+T_{\rm IS}^2}
  \right)\,
  \Phi_{\bar p}^{\rm IS}(T_{\rm IS}),
\end{equation}
where $T_{\rm IS}=T_{\rm TOA}+\phi_F$, with $T_{\rm IS}$ and $T_{\rm TOA}$
being the antiproton kinetic energies at the heliospheric boundary and at the
top of the Earth's atmosphere, respectively.

\subsection{Gamma rays from inverse Compton scattering}
As discussed above, electrons and positrons from dark matter decay lose their
energy mainly via interaction with the Galactic magnetic field and the ISRF.
In the first case (assuming injection energies of the order of $1\TeV$)
synchrotron radiation in the radio band with frequencies $\mathcal{O}(0.1 -
100~\text{GHz})$ is produced and potentially observable (see {\it e.g.}
Ref.~\cite{synchrotron}). In the second case, the ICS of electrons and
positrons on the ISRF (which includes the cosmic microwave background, thermal
dust radiation and starlight) produces gamma rays with energies between
$100\MeV$ and $1\TeV$. Recently, ICS in connection with the PAMELA excess was
discussed in Refs.~\cite{Cirelli:2009vg, Ishiwata:2009dk}, and we will follow
their treatment. A pedagogical review of ICS can be found in
Ref.~\cite{Blumenthal:1970gc}.

The rate of inverse Compton scattering of an electron with energy $E_e$, where
an ISRF photon with an energy between $\epsilon$ and $\epsilon+d\epsilon$ is
upscattered to energies between $E_\gamma$ and $E_\gamma+dE_\gamma$, is given
by
\begin{eqnarray}
  \frac{dN(E_e,\vec{r})}{d\epsilon\,dE_\gamma\,dt}=
  \frac{3}{4}\frac{\sigma_\text{T}}{\gamma^2\, \epsilon}\,
  f_\text{ISRF}(\epsilon,\vec{r})\,
  \left[
  2q\ln q + 1 + q - 2q^2
  +\frac{1}{2}\frac{(q\Gamma)^2}{1+q\Gamma}(1-q)\right],
  \label{eqn:ICrate}
\end{eqnarray}
where $\sigma_\text{T}=0.67\;\text{barn}$ denotes the Compton scattering cross
section in the Thomson limit, $\gamma=E_e/m_e$, $q=E_\gamma m_e
/(4\epsilon\gamma(m_e\gamma-E_\gamma))$, and $f_\text{ISRF}(\epsilon,\vec{r})$
denotes the differential number density of ISRF photons with energy
$\epsilon$, at spatial position $\vec{r}$.

From Eq.~\eqref{eqn:ICrate} one can derive the energy loss
$b_{\text{ICS}}(E_e,\vec{r})$ of electrons due to ICS (which represents the
dominant contribution to the energy loss rate $b(E_e,\vec{r})$ in the
transport equation, Eq. (\ref{transport1})), and the corresponding power
$\mathcal{P}(E_\gamma,E_e,\vec{r})$ that is emitted in gamma rays with
energies between $E_\gamma$ and $E_\gamma+dE_\gamma$ via
\begin{eqnarray}
  b_{\text{ICS}}(E_e,\vec{r})=\int_0^\infty d\epsilon
  \int_{\sim\epsilon}^{4\epsilon\gamma^2} dE_\gamma (E_\gamma-\epsilon)
  \frac{dN(E_e,\vec{r})}{d\epsilon\,dE_\gamma\,dt} 
\end{eqnarray}
and
\begin{eqnarray}
  \mathcal{P}(E_\gamma,E_e,\vec{r})=\int_0^\infty d\epsilon
  (E_\gamma-\epsilon)
  \frac{dN(E_e,\vec{r})}{d\epsilon\,dE_\gamma\,dt}.
\end{eqnarray}

Neglecting diffusion and synchrotron losses\footnote{In Ref.~\cite{Meade} it
was discussed that this approximation gives results for the ICS fluxes that
are correct at the $\mathcal{O}(2)$ level, which is sufficient for our
analysis.}, the energy distribution of electrons and positrons from dark
matter decay it is given by
\begin{eqnarray}
  f_{e^\pm}(E_e,\vec{r})=\frac{1}{b_{\text{ICS}}(E_e,\vec{r})}
  \int_{E_e}^{M_\text{DM}} d\tilde{E}\; Q_{e^\pm}(\tilde{E},\vec{r}).
  \label{eqn:DiffAp}
\end{eqnarray}

For the differential flux of ICS photons of energy $E_\gamma$ from a region
$\Delta\Omega$ of the sky it then follows:
\begin{eqnarray}
  \frac{d\Phi}{dE_\gamma}=2\,
  \frac{1}{4\pi\, E_\gamma\, \tau_\text{DM}}
  \int_{\Delta\Omega} d\Omega \int_\text{l.o.s.}ds\,
  \frac{\rho(\vec{r})}{M_\text{DM}}
  \int_{m_e}^{M_\text{DM}} dE_e\;
  \frac{\mathcal{P}(E_\gamma,E_e,\vec{r})}{b_{\text{ICS}}(E_e,\vec{r})}\,
  Y(E_e),
  \label{}
\end{eqnarray}
where $Y(E_e)=\int_{E_e}^{M_\text{DM}} d\tilde{E}\, dN_{e^\pm}/d\tilde{E}$
describes the number of particles in the spectrum of electrons and positrons
above a certain energy $E_e$. In the second integral, the coordinate $s$ runs
over the line of sight (l.o.s.), which points in the direction of
$\Delta\Omega$. The prefactor $2$ takes into account that the same amount of
gamma rays comes from the dark matter electrons and positrons. In this work,
we use the ISRF data as derived in Ref.~\cite{Porter:2005qx}, and we fully
take into account the spatial dependence of the energy loss
$b_{\text{ICS}}(E_e,\vec{r})$ in Eq.~\eqref{eqn:DiffAp}. Furthermore, we
calculate the gamma rays from ICS with extragalactic origin. In this case,
effects of redshifting must also be taken into account. Details of this
calculation can be found in Ref.~\cite{Ishiwata:2009dk}. 

In addition to the gamma-ray signal from dark matter decay there exists a
background contribution, presumably originating from active galactic nuclei
(AGN), which is perfectly isotropic, and which has an energy spectrum which is
assumed to follow a simple power law; the normalization and index will be
treated as free parameters to be determined by requiring a good fit of the
total flux to the data.

We will compare our predicted flux to two sets of data for the extragalactic
diffuse gamma-ray background obtained from the EGRET data, using two different
models for the Galactic background, averaging over the whole sky, excluding
the region of the Galactic plane with latitudes $|b|<10^\circ$. The first
analysis of the extragalactic diffuse gamma-ray flux by Sreekumar et
al.~\cite{sxx97} revealed a power law
\begin{equation}
  \left[E^2 \frac{dJ}{dE}\right]_\text{bkg}=1.37\times 10^{-6}
  \left(\frac{E}{\rm GeV}\right)^{-0.1}
  (\text{cm}^2~\text{str}~\text{s})^{-1}~\text{GeV} 
\end{equation}
in the energy range $50\;{\rm MeV}-10\;{\rm GeV}$. On the other hand, the
extraction of the extragalactic background by Strong, Moskalenko and Reimer
\cite{smr04}, using an optimized model to better simulate the Galactic diffuse
emission, revealed a steeper power law,
\begin{equation}
  \left[E^2 \frac{dJ}{dE}\right]_\text{bkg}=6.8\times 10^{-7}
  \left(\frac{E}{\rm GeV}\right)^{-0.32}
  (\text{cm}^2~\text{str}~\text{s})^{-1}~\text{GeV},
\end{equation}
between $50\;{\rm MeV}-2\;{\rm GeV}$ and an intriguing break of the spectrum
at energies $2\;{\rm GeV}-10\;{\rm GeV}$. Future measurements by the Fermi
LAT, as well as a better understanding of the Galactic diffuse emission, will
provide a better determination of the extragalactic diffuse gamma-ray flux in
the near future.

\section{Predictions from decaying dark matter}
We will analyze in this section the predictions for the positron fraction and
the total electron-plus-positron flux including a possible contribution from
dark matter decay in order to account for the anomalies observed by PAMELA and
Fermi. To keep the analysis as model-independent as possible, we will analyze
several scenarios of decaying dark matter, computing the predictions for the
positron fraction and the total electron-plus-positron flux for either a
fermionic or a bosonic particle, which decays into various channels with a
branching ratio of 100\%. We calculated for each of these channels the energy
spectrum of electrons and positrons using the event generator PYTHIA
6.4~\cite{SMS06}. Thus, from the particle physics point of view the only free
parameters are the dark matter mass and lifetime. From the astrophysics point
of view there are a number of uncertainties, such as the choice of propagation
parameters and the choice of the background fluxes of electrons and positrons.
As mentioned in the previous section, we will adopt the MED propagation model
defined in \cite{MDT+01}, which provides the best fit to the Boron-to-Carbon
(B/C) ratio, although the results are not very sensitive to the particular
choice of the propagation model. On the other hand, for the background fluxes
of electrons and positrons we will adopt the spectra corresponding to the
``model 0'' proposed by the Fermi collaboration. However, we will allow for a
possible shift in the normalization of the background flux of electrons, which
is dominated by primaries, due to our ignorance of the amount of electrons
injected in the interstellar medium. In our analysis we will sample several
dark matter masses and treat the dark matter lifetime and the normalization of
the background flux of electrons as free parameters which will be determined
to provide a qualitatively good fit to the PAMELA and Fermi measurements. Note
that below energies of $10\GeV$ the data is best fitted for normalizations
$k\simeq1$. In our plots, we always used normalization factors $k\geq 0.8$.

Let us now discuss the cases of fermionic and scalar dark matter particles
separately. 

\subsection{Fermionic dark matter decay}
In the case where the dark matter particle is a fermion $\psi_\text{DM}$, we
consider the following decay channels\footnote{We do not include quarks or
Higgs bosons in the list, since they yield similar signatures to gauge boson
fragmentation. Furthermore, we only consider decay channels with two or three
final-state particles.}:
\begin{eqnarray}
  \psi_\text{DM}&\rightarrow& Z^0 \nu\;, \nonumber \\ 
  \psi_\text{DM}&\rightarrow& W^\pm \ell^\mp\;, \nonumber\\
  \psi_\text{DM}&\rightarrow& \ell^+ \ell^- \nu\;,
\end{eqnarray}
where the three-body decay into charged leptons and a neutrino is assumed to
be mediated by the exchange of a scalar particle, motivated by the interesting
scenario of a hidden gaugino as dark matter particle~\cite{gaugino}.

\begin{figure}[ht]
  \begin{center}
    \includegraphics[width=.95\linewidth]{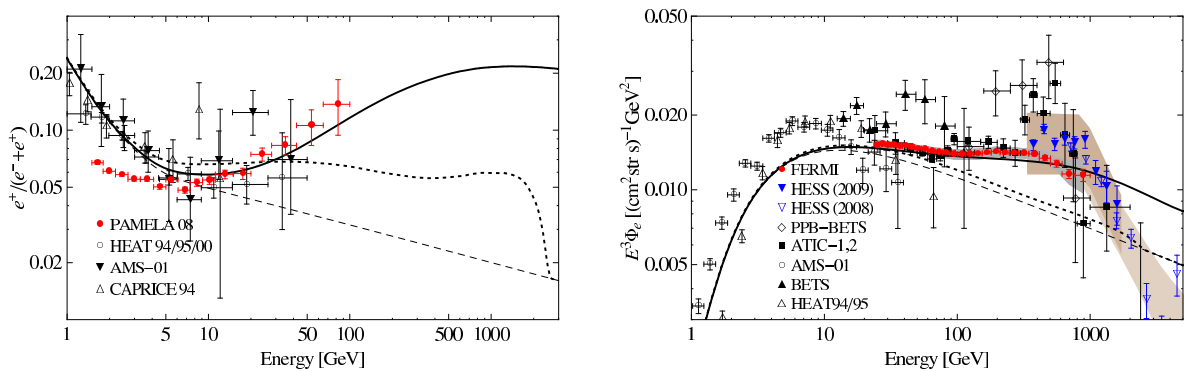}
    \vspace{-1cm}
  \end{center}
  \caption{Positron fraction ({\it left panel}) and total
  electron-plus-positron flux ({\it right panel}) for the decay channel
  $\psi_\text{DM}\rightarrow Z^0 \nu$ with $M_\text{DM}=100\TeV$ (solid) and
  $5\TeV$ (dotted). The dashed line shows the background fluxes as discussed
  in the text. Solar modulation is taken into account using the force field
  approximation with $\phi_F=550\,\text{MV}$.}
  \label{fig:Znu}
\end{figure}

\begin{figure}[ht]
  \begin{center}
    \includegraphics[width=.95\linewidth]{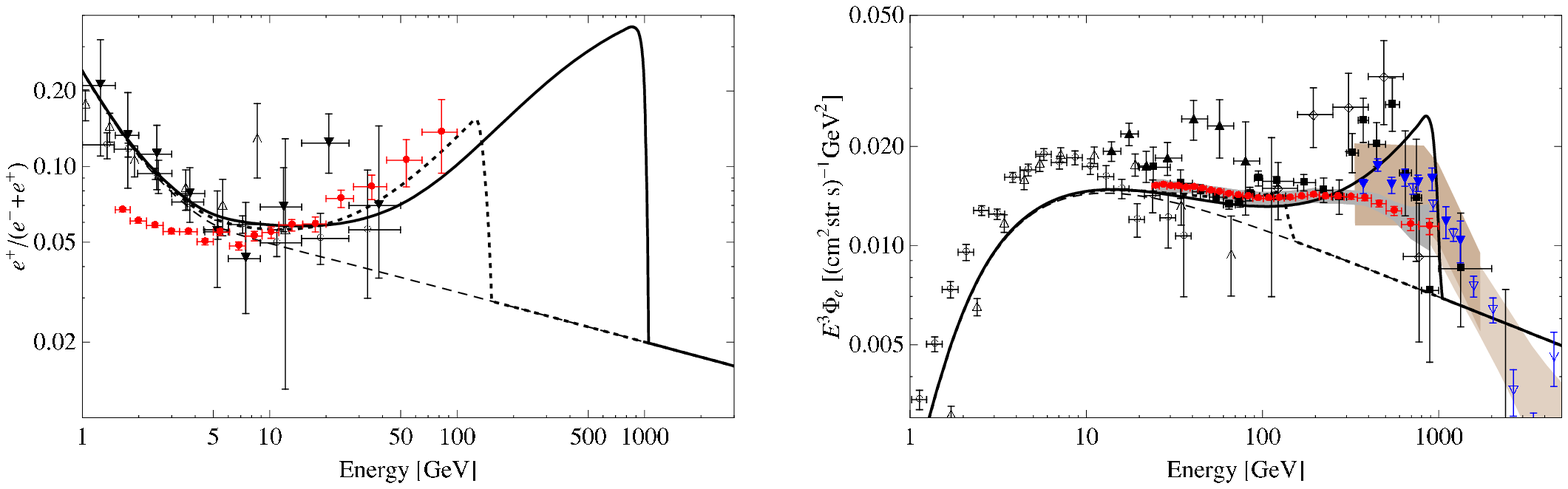}
    \includegraphics[width=.95\linewidth]{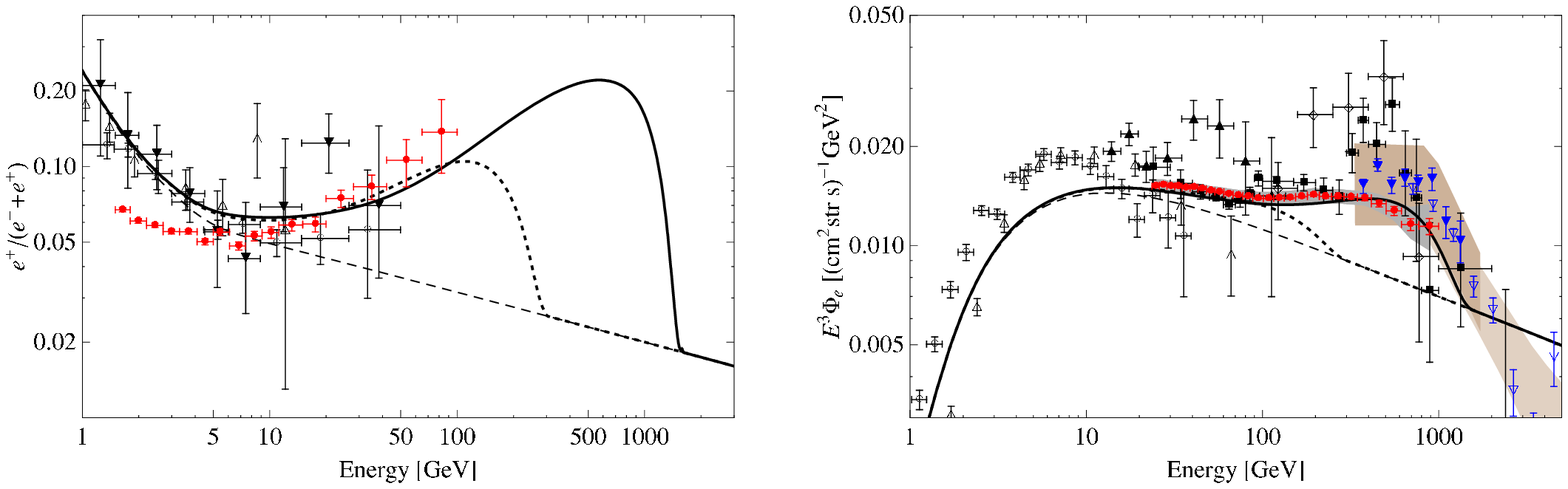}
    \includegraphics[width=.95\linewidth]{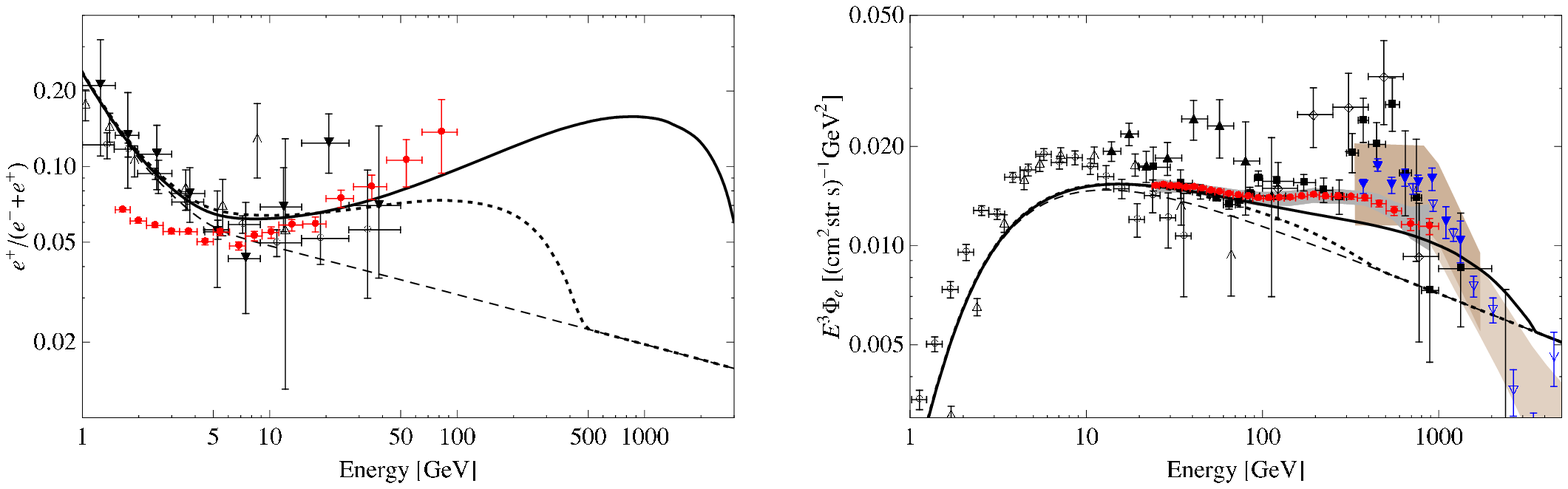}
    \vspace{-1cm}
  \end{center}
  \caption{Same as Fig.~\ref{fig:Znu}, but for the decay channels
  $\psi_\text{DM}\rightarrow W^\pm \ell^\mp$. {\it Upper panels:}
  $\psi_\text{DM}\rightarrow W^\pm e^\mp$ with $M_\text{DM}=2000\GeV$ (solid)
  and $300\GeV$ (dotted). {\it Middle panels:} $\psi_\text{DM}\rightarrow
  W^\pm\mu^\mp$ with $M_\text{DM}=3000\GeV$ (solid) and $600\GeV$ (dotted).
  {\it Lower panels:} $\psi_\text{DM}\rightarrow W^\pm\tau^\mp$ with
  $M_\text{DM}=8000\GeV$ (solid) and $1000\GeV$ (dotted).}
  \label{fig:3Wl}
\end{figure}

\begin{figure}[ht]
  \begin{center}
    \includegraphics[width=.45\linewidth]{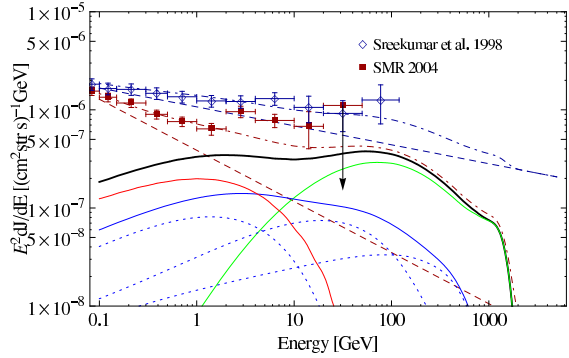}
    \vspace{-1cm}
  \end{center}
  \caption{Extragalactic diffuse gamma-ray flux for $\psi_\text{DM}\rightarrow
  W^\pm\mu^\mp$ with $M_\text{DM}=3000\GeV$ and
  $\tau_\text{DM}=2.1\times10^{26}\s$. The gamma-ray flux is averaged over the
  whole sky, excluding the Galactic plane, $|b|<10^\circ$. We included gamma
  rays produced directly in the final-state radiation of the muons and the
  fragmentation of $W^\pm$ (green line), gamma rays from ICS of dark matter
  electrons and positrons on the ISRF (solid blue line; the dotted blue lines
  show, from left to right, the fluxes that come from scattering on the CMB,
  on the thermal radiation of dust and on starlight) and gamma rays from ICS
  outside of our Galaxy (red). The black solid line shows the overall flux.
  The dark red and dark blue lines show the total flux (dash-dotted) adding an
  isotropic extragalactic background (dashed) with a power-law spectrum.
  Normalization and power index are chosen to fit one of the two data sets
  shown \cite{sxx97,smr04}.}
  \label{fig:Wmu_addon}
\end{figure}

\begin{figure}[ht]
  \begin{center}
    \includegraphics[width=.45\linewidth]{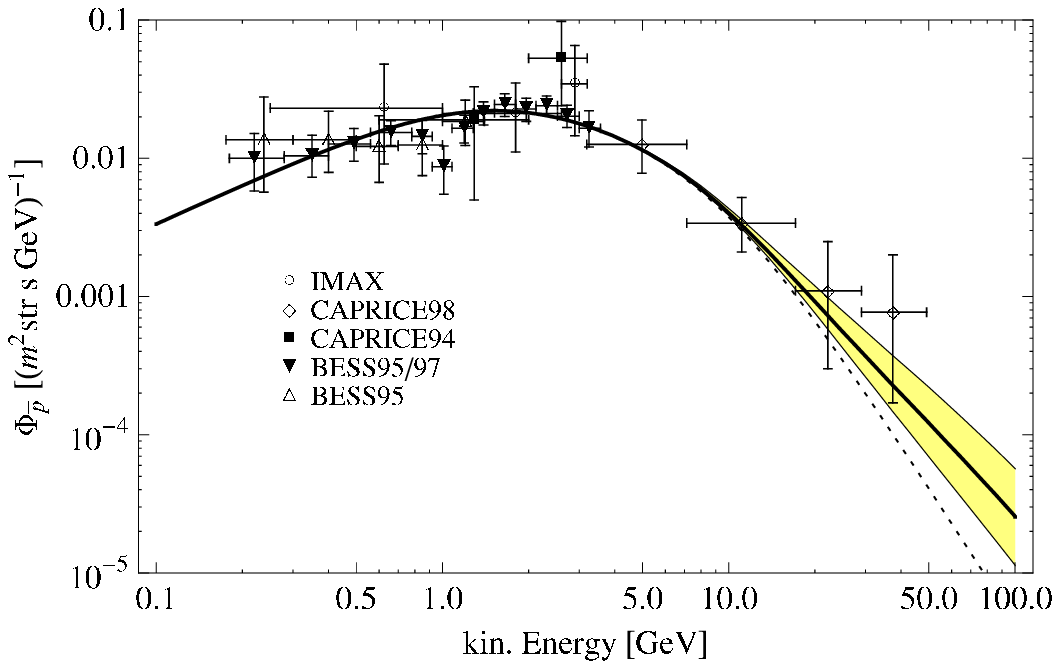}
    \includegraphics[width=.45\linewidth]{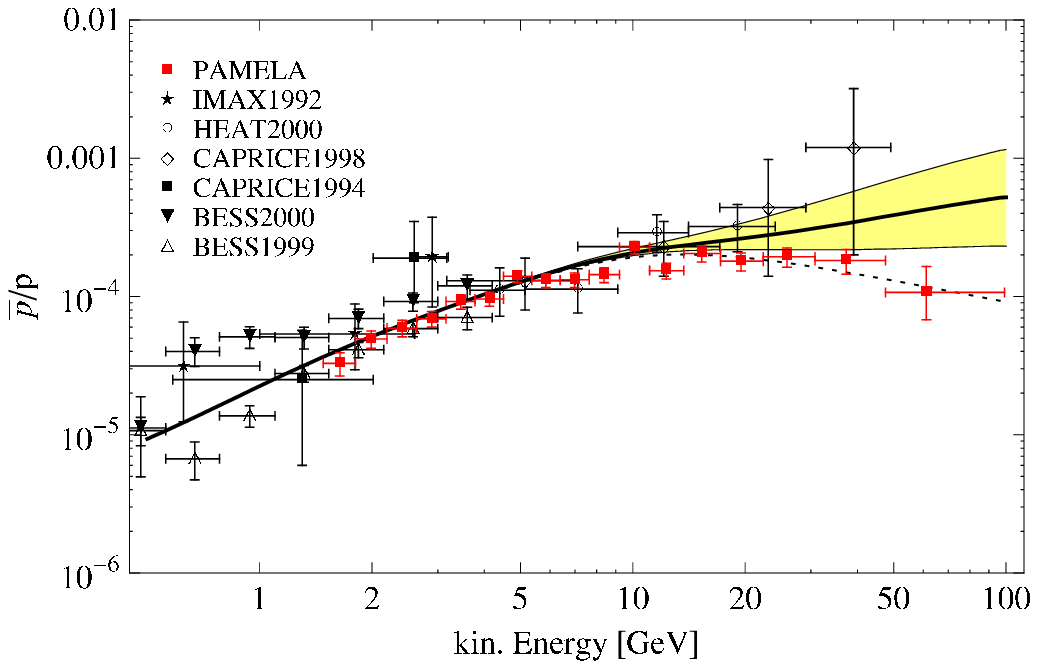}
    \vspace{-1cm}
  \end{center}
  \caption{Antiproton flux ({\it left panel}) and the corresponding
  antiproton-to-proton ratio ({\it right panel}) for
  $\psi_\text{DM}\rightarrow W^\pm\mu^\mp$ with $M_\text{DM}=3000\GeV$ and
  $\tau_\text{DM}=2.1\times10^{26}\s$. For the antiproton flux we adopt the
  background from Ref.~\cite{Bringmann:2006im}, while antiproton-to-proton
  ratio is plotted using the background from Ref.~\cite{Lionetto:2005jd}, and
  the yellow band indicates the uncertainties from the propagation model.  The
  solid black line corresponds to the MED model of
  Tab.~\ref{tab:param-antiproton}.}
  \label{fig:Wmu_pbar}
\end{figure}

\begin{figure}[ht]
  \begin{center}
    \includegraphics[width=.95\linewidth]{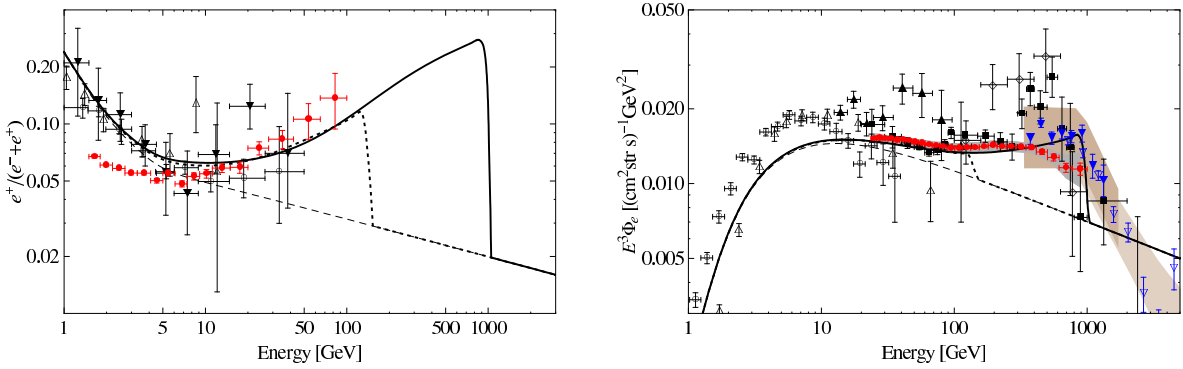}
    \vspace{-1cm}
  \end{center}
  \caption{Same as Fig.~\ref{fig:Znu}, but for the flavor-democratic decay
  $\psi_\text{DM}\rightarrow W^\pm \ell^\mp$ with equal branching ratios into
  the three charged lepton flavors, for $M_\text{DM}=2000\GeV$ (solid) and
  $300\GeV$ (dotted).}
  \label{fig:Wlep}
\end{figure}

\begin{figure}[ht]
  \begin{center}
    \includegraphics[width=.95\linewidth]{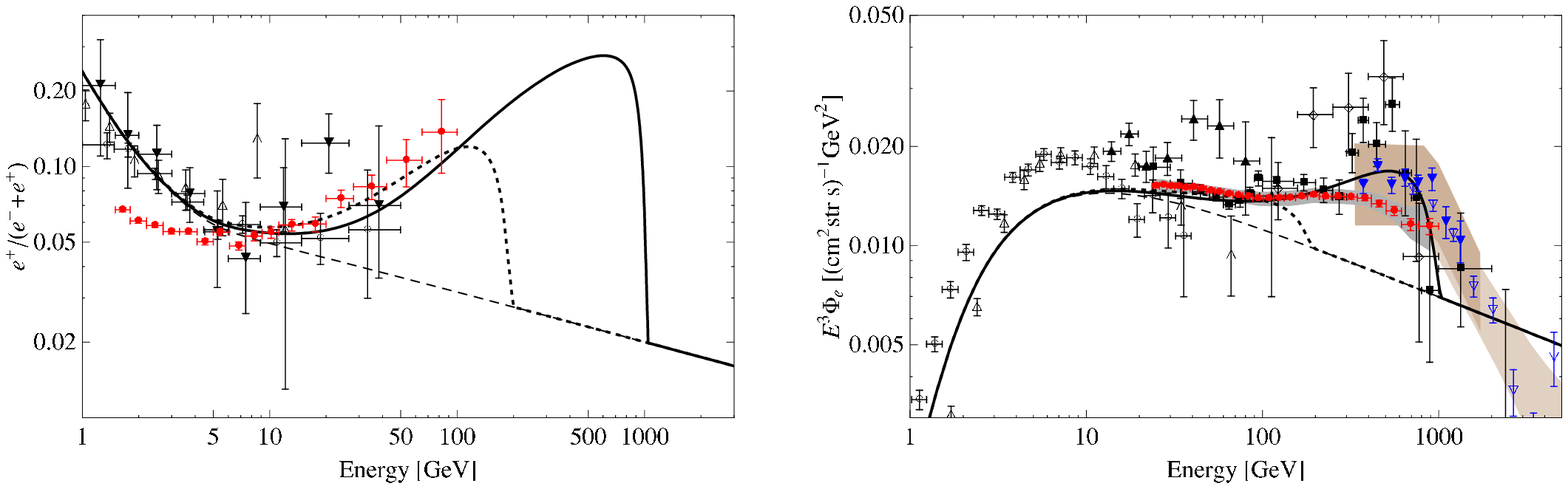}
    \includegraphics[width=.95\linewidth]{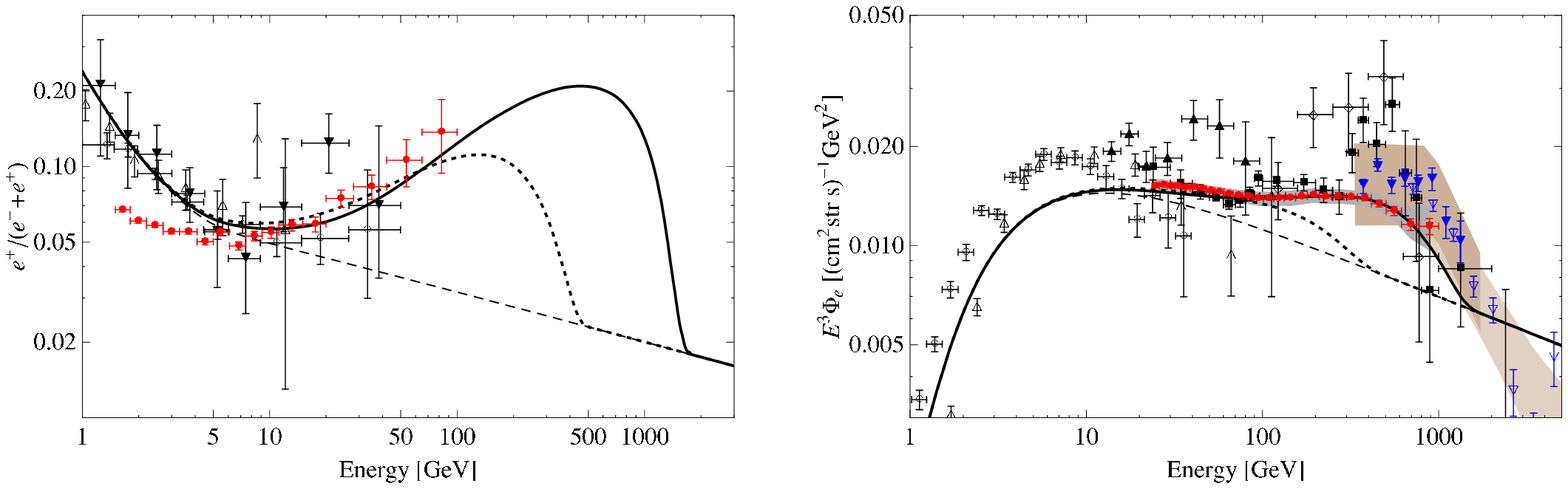}
    \includegraphics[width=.95\linewidth]{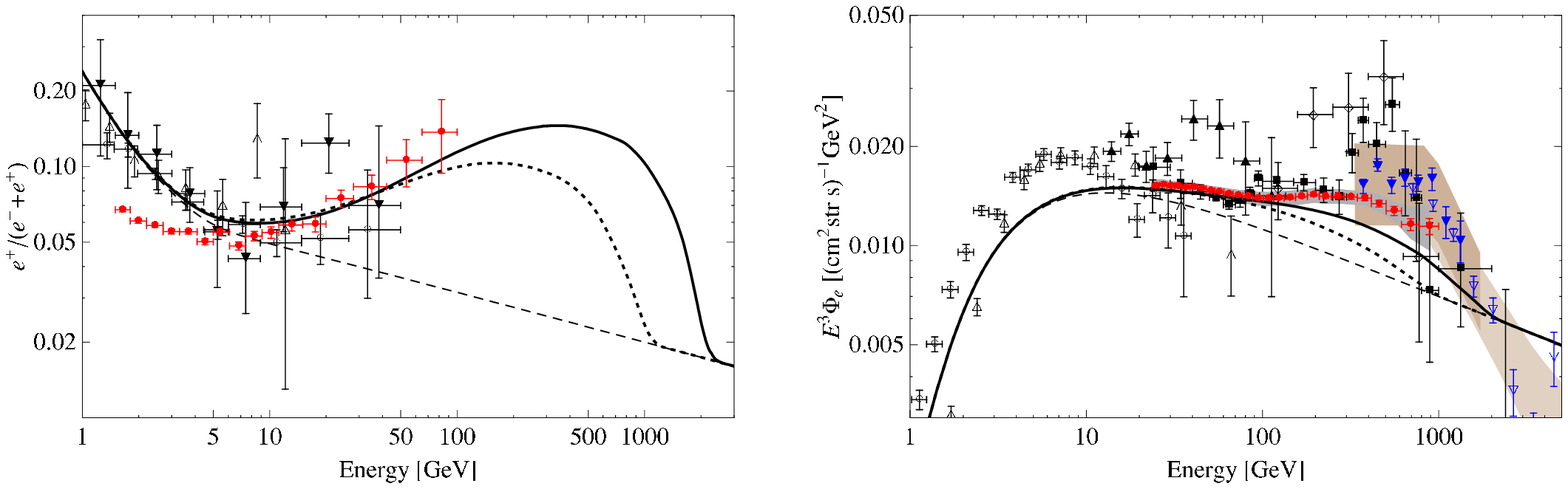}
    \vspace{-1cm}
  \end{center}
  \caption{Same as Fig.~\ref{fig:Znu}, but for the decay channels
  $\psi_\text{DM}\rightarrow \ell^\pm \ell^\mp \nu$. {\it Upper panels:}
  $\psi_\text{DM}\rightarrow e^-e^+\nu$ with $M_\text{DM}=2000\GeV$ (solid)
  and $400\GeV$ (dotted). {\it Middle panels:} $\psi_\text{DM}\rightarrow
  \mu^-\mu^+\nu$ with $M_\text{DM}=3500\GeV$ (solid) and $1000\GeV$ (dotted).
  {\it Lower panels:} $\psi_\text{DM}\rightarrow \tau^-\tau^+\nu$ with
  $M_\text{DM}=5000\GeV$ (solid) and $2500\GeV$ (dotted).}
  \label{fig:3llnu}
\end{figure}

\begin{figure}[ht]
  \begin{center}
    \includegraphics[width=.95\linewidth]{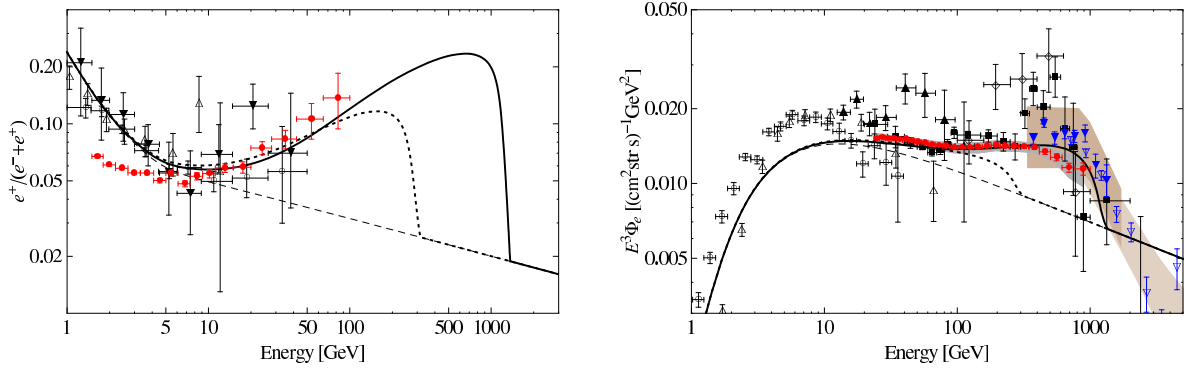}
    \vspace{-1cm}
  \end{center}
  \caption{Same as Fig.~\ref{fig:Znu}, but for the democratic decay
  $\psi_\text{DM}\rightarrow \ell^\pm \ell^\mp \nu$ with equal branching
  ratios into the three charged lepton flavors, with $M_\text{DM}=600\GeV$
  (dotted) and $2500\GeV$ (solid).}
  \label{fig:leplepnu}
\end{figure}

\begin{figure}[ht]
  \begin{center}
    \includegraphics[width=.45\linewidth]{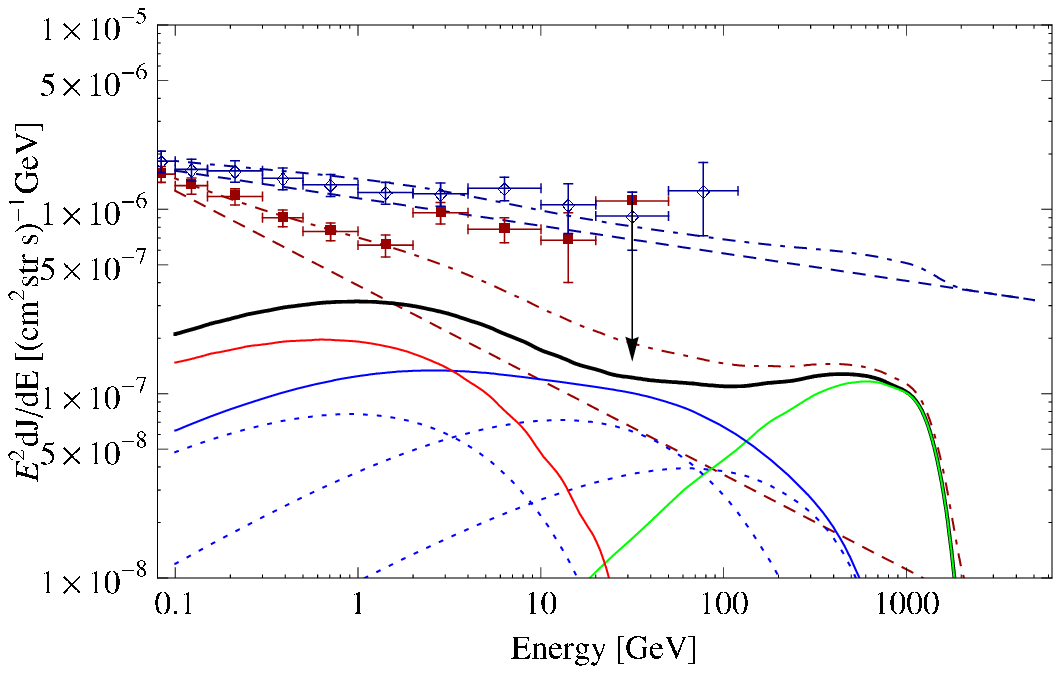}
    \includegraphics[width=.45\linewidth]{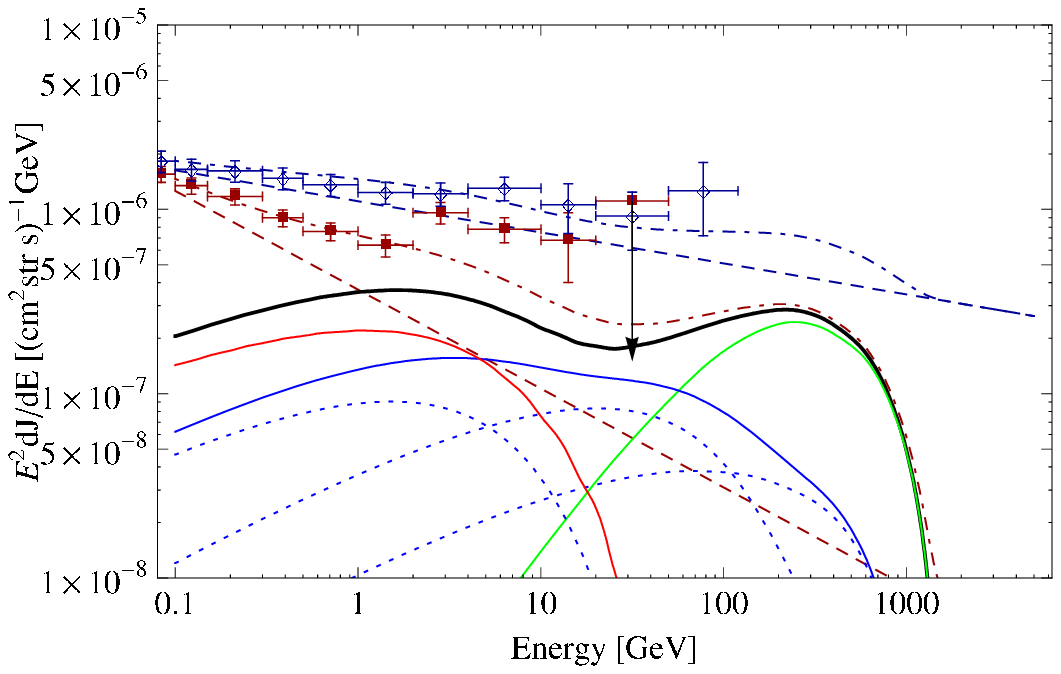}
    \vspace{-1cm}
  \end{center}
  \caption{Same as Fig.~\ref{fig:Wmu_addon}, but for
  $\psi_\text{DM}\rightarrow\mu^-\mu^+\nu$ ({\it left panel}, with
  $M_\text{DM}=3500\GeV$) and for the democratic decay
  $\psi_\text{DM}\rightarrow \ell^-\ell^+\nu$ ({\it right panel}, with
  $M_\text{DM}=2500\GeV$).}
  \label{fig:mumunu_addon}
\end{figure}

The predicted positron fraction in the case where the dark matter particles
decay via $\psi_\text{DM} \rightarrow Z^0 \nu$ is shown in the left panel of
Fig.~\ref{fig:Znu}, compared to the PAMELA, HEAT, CAPRICE and AMS-01 data, for
the exemplary dark matter masses $M_{\rm DM}=5$ and $100\TeV$. In the right
panel we show the corresponding total electron-plus-positron flux compared to
the results from Fermi, H.E.S.S., PPB-BETS, BETS, ATIC, HEAT, CAPRICE and
AMS-01. The dark matter lifetimes and the normalization factors $k$ of the
primary electron flux have been chosen in each case to provide a reasonable
fit to the PAMELA and Fermi data points. In this decay channel, the dominant
source of electrons and positrons is the fragmentation of the $Z^0$ boson
(with a rather small branching ratio of $Z^0$ decays into a pair of charged
leptons), which produces relatively soft particles. As a result, even though
this decay mode can produce a visible excess in the positron fraction, the
energy spectrum is in general too flat to explain the steep rise observed by
PAMELA. An exception occurs if the dark matter mass is very large,
$M_\text{DM}\gtrsim 50\TeV$. In this case, the electrons and positrons from
dark matter decay are boosted to high enough energies to produce the steep
rise in the positron fraction. However, these large dark matter masses seem to
be in conflict with the H.E.S.S. observations, which require a falloff in the
total electron-plus-positron spectrum at $\sim1\TeV$.

On the other hand, we show in Fig.~\ref{fig:3Wl} the predictions for the
cosmic-ray electron and positron fluxes when a fermionic dark matter particle
decays as $\psi_\text{DM}\rightarrow W^\pm \ell^\mp$ for different dark matter
masses. The electrons and positrons created in the fragmentation of the
$W^\pm$ gauge bosons produce a rather flat contribution to the positron
fraction. However, the hard electrons and positrons resulting from the decay
of the $\mu^\pm$ and $\tau^\pm$ leptons or directly from the dark matter decay
into $e^\pm$ produce a rise in the total energy spectrum and in the positron
fraction. The decay mode $\psi_\text{DM}\rightarrow W^\pm e^\mp$, which can
produce a steep rise in the positron fraction and is thus consistent with the
PAMELA observations, produces also a steep rise and a sharp falloff in the
total electron-plus-positron flux, which is not observed by Fermi. Thus, the
possibility that the dark matter particles decay preferentially in this decay
mode, which is well compatible the PAMELA observations, is now strongly
disfavored by the Fermi results on the total electron-plus-positron flux. 

The decay mode $\psi_\text{DM}\rightarrow W^\pm \mu^\mp$, however, can nicely
accommodate the PAMELA and Fermi observations when the dark matter mass is
$M_\text{DM}\simeq3\TeV$ and the lifetime is
$\tau_\text{DM}\simeq2.1\times10^{26}\s$. In this decay mode, the
fragmentation of the $W^\pm$ gauge bosons also produces fluxes of primary
antiprotons and gamma rays, which are severely constrained by present
experiments. The predictions for the gamma-ray and antiproton fluxes for this
particular decay mode are shown in Figs.~\ref{fig:Wmu_addon} and
\ref{fig:Wmu_pbar}; the former figure shows the gamma-ray fluxes from
final-state radiation and $W^\pm$ fragmentation (green), and from Galactic
(blue) and extragalactic (red) ICS of dark matter electrons and positrons. We
also show the total flux compared to the extraction of the extragalactic
diffuse gamma-ray flux by Sreekumar et al.~\cite{sxx97} and by Strong,
Moskalenko and Reimer \cite{smr04}, averaging over the whole sky excluding the
region of the Galactic plane with latitudes $|b|<10^\circ$ and assuming a
power law for the genuinely extragalactic component. On the other hand, the
latter figure shows the prediction for the antiproton-to-proton ratio with an
uncertainty band corresponding to the MAX, MED and MIN models in
Tab.~\ref{tab:param-antiproton}.  While the absolute flux is compatible with
existing measurements, it is apparent from the figure that the
antiproton-to-proton ratio is in some tension with the results at the highest
energies explored by PAMELA. The fragmentation of the $W^\pm$ gauge bosons
also produces a sizable contribution to the total gamma-ray flux at high
energies which could be visible by the Fermi LAT as a bump over the
background, which is assumed to follow a simple power law, especially if it
has a large index as in the extraction of the diffuse extragalactic background
from the EGRET data by Strong, Moskalenko and Reimer.  Lastly, the decay mode
$\psi_\text{DM}\rightarrow W^\pm \tau^\mp$ predicts, for a wide range of dark
matter masses, a positron fraction and an electron-plus-positron flux that are
too flat to explain the anomalies observed by PAMELA and Fermi. 

In some decaying dark matter scenarios, the dark matter particles decay into
charged leptons of different flavors and not exclusively in just one channel.
As an illustration of the predictions of this class of scenarios, we show in
Fig.~\ref{fig:Wlep} the positron fraction and the total electron-plus-positron
flux for a dark matter particle that decays democratically into the three
flavors, for $M_\text{DM}=2000\GeV$ (solid) and $300\GeV$ (dotted). Although
these scenarios could explain the PAMELA excess, the predicted spectral shape
of the total flux is not consistent with the Fermi data: either the energy
spectrum falls off at too low energies or it presents a sharp peak at high
energies, due to the large branching ratio into hard electrons and positrons.
Scenarios with smaller branching ratio into electron flavor and larger
branching ratio into muon flavor could, however, explain both anomalies
simultaneously.

The dark matter particles could also decay into three fermions, namely into a
lepton-antilepton pair and a neutrino. In this case many possibilities could
arise depending on the specific particle physics scenario. We will just
concentrate on the case where the lepton and the antilepton carry the same
flavor and the decay is mediated by a heavy scalar\footnote{Our results are
not very sensitive to the mass splitting between dark matter particle and
virtual scalar.}. The results for the positron fraction and the total
electron-plus-positron flux are shown in Fig.~\ref{fig:3llnu}. 

The spectrum produced in the decay into electron-positron pairs is flatter in
this case than in the two-body decay $\psi_\text{DM}\rightarrow W^\pm e^\mp$,
although it still predicts a rather prominent bump in the electron spectrum at
high energies, which is not observed by Fermi. More promising is the decay
channel $\psi_\text{DM}\rightarrow \mu^-\mu^+\nu$, which can reproduce quite
nicely the Fermi electron-plus-positron spectrum and the steep rise in the
positron fraction observed by PAMELA when the dark matter mass is
$M_\text{DM}\simeq3500\GeV$ and the lifetime is
$\tau_\text{DM}\simeq1.1\times10^{26}\s$. Lastly, decays into tau flavor can
qualitatively reproduce the steep rise in the positron fraction for dark
matter masses above $\sim 2.5\TeV$, although as apparent from
Fig.~\ref{fig:3llnu}, lower right panel, the resulting electron-plus-positron
spectrum has an energy dependence much steeper than $E^{-3.0}$ at high
energies, in tension with the Fermi measurements. In this case an additional
source of high-energy positrons, coming {\it e.g.} from pulsars, must be
invoked in order to reproduce the Fermi energy spectrum.

As for the two-body decays $\psi_\text{DM}\rightarrow W^\pm \ell^\mp$, the
dark matter particle could also decay into charged fermions with different
flavor. We illustrate such a situation showing in Fig.~\ref{fig:leplepnu} the
predictions for the positron fraction and the total electron-plus-positron
flux when the dark matter particles decay democratically into the three
flavors, for dark matter masses $M_\text{DM}=600\GeV$ (dotted) and $2500\GeV$
(solid). In particular, this is the case in scenarios where dark matter
neutralinos decay into light hidden gauginos via kinetic mixing, or vice
versa~\cite{gaugino}. It is interesting that these scenarios can
simultaneously explain the PAMELA and Fermi anomalies when the dark matter
mass is $M_\text{DM} \simeq 2500\GeV$. For the two cases of three-body decay
into muon flavors and democratic decay, we show the predictions for the
extragalactic diffuse gamma-ray fluxes in Fig.~\ref{fig:mumunu_addon}. In both
cases, they are consistent with the present data and show a deviation from the
putative power-law behavior of the astrophysical background, which could be
observed by the Fermi LAT, depending on the precise spectrum of the genuinely
extragalactic contribution to the flux.

\begin{figure}[ht]
  \begin{center}
    \includegraphics[width=.95\linewidth]{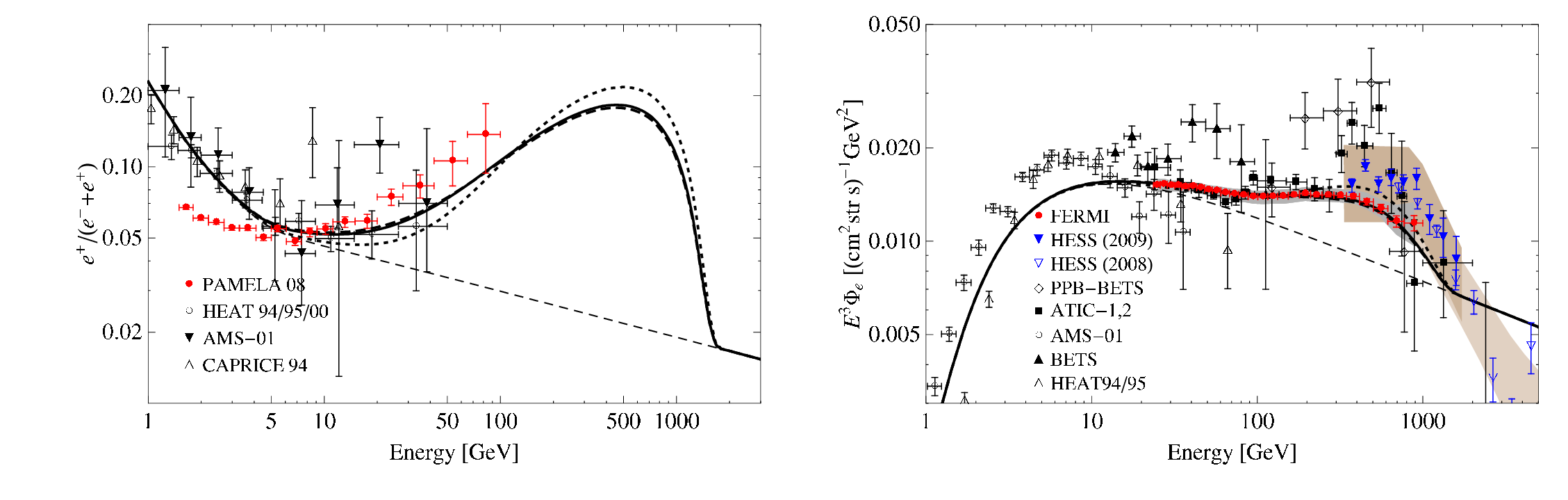}
    \vspace{-1cm}
  \end{center}
  \caption{Illustration of the dependence on the choice of transport
  parameters. Same as Fig. \ref{fig:3llnu}, middle panels, but only for a dark
  matter mass of 3500 GeV. The solid, dashed and dotted lines correspond to
  the MED, MAX and MIN model parameters, respectively. The results for the MED
  and MAX model are very similar because the height of the diffusion zone
  becomes irrelevant above a few kpc for high-energy electrons from local
  sources.}
  \label{fig:MINMEDMAX}
\end{figure}

We summarize our results for the promising fermionic dark matter scenarios,
together with the corresponding dark matter masses and lifetimes, in
Tab.~\ref{tab:results}. The impact of choosing other sets of propagation
parameters is illustrated in Fig. \ref{fig:MINMEDMAX} for the decay mode
$\psi_{\text{DM}} \rightarrow \mu^+\mu^-\nu$.

\subsection{Scalar dark matter decay}
For a scalar dark matter particle, we will discuss the following decay
channels\footnote{Again, we do not include quarks or Higgs bosons in the list.
Three-body decay modes like $\phi_\text{DM}\rightarrow \ell^+\ell^-\gamma$ are
expected to give results similar to the fermionic dark matter case.}:
\begin{eqnarray}
  \phi_\text{DM}&\rightarrow& Z^0 Z^0, \nonumber \\ 
  \phi_\text{DM}&\rightarrow& W^+ W^-,  \nonumber\\
  \phi_\text{DM}&\rightarrow& \ell^+ \ell^- .
\end{eqnarray}

\begin{figure}[ht]
  \begin{center}
    \includegraphics[width=.95\linewidth]{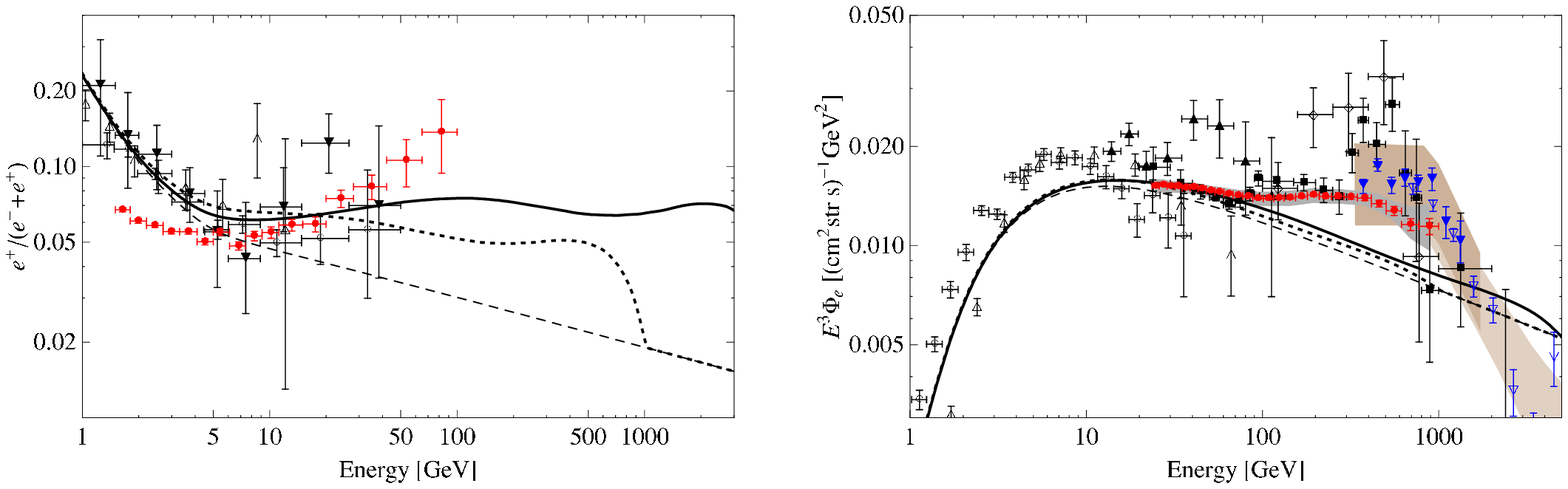}
    \vspace{-1cm}
  \end{center}
  \caption{Same as Fig.~\ref{fig:Znu}, but for the decay channel
  $\phi_\text{DM}\rightarrow Z^0Z^0$ with $M_\text{DM}=10\TeV$ (solid) and
  $2\TeV$ (dotted).}
  \label{fig:ZZ}
\end{figure}

\begin{figure}[ht]
  \begin{center}
    \includegraphics[width=.95\linewidth]{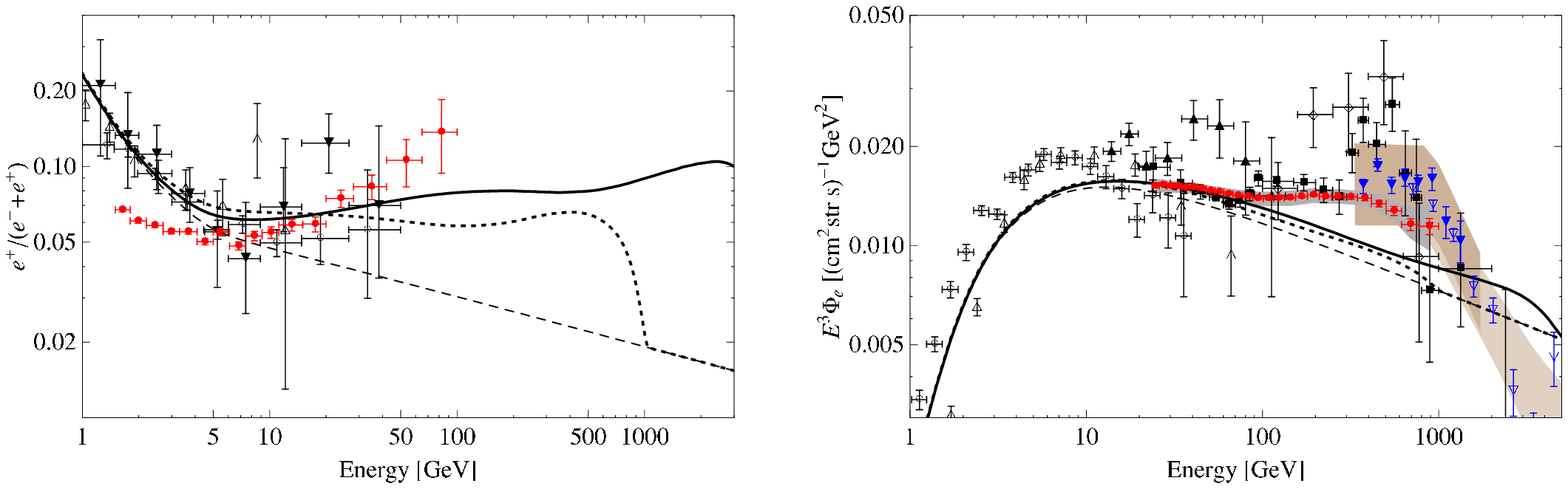}
    \vspace{-1cm}
  \end{center} 
  \caption{Same as Fig.~\ref{fig:Znu}, but for the decay channel
  $\phi_\text{DM}\rightarrow W^+W^-$ with $M_\text{DM}=10\TeV$ (solid) and
  $2\TeV$ (dotted).}
  \label{fig:WW}
\end{figure}

We show in Figs.~\ref{fig:ZZ} and \ref{fig:WW} the positron fraction and the
total electron-plus-positron flux for a scalar dark matter particle that
decays exclusively into weak gauge bosons $\phi_\text{DM} \rightarrow Z^0 Z^0$
and $\phi_\text{DM} \rightarrow W^+ W^-$ for dark matter masses $M_{\rm
DM}=2\TeV$ and $10\TeV$. As generically expected from decays into weak gauge
bosons, the electrons and positrons produced are relatively soft, resulting in
a positron fraction which is too flat to explain the steep rise in the
spectrum observed by PAMELA (for an exception with large dark matter masses
see Fig.~\ref{fig:Znu}).

\begin{figure}[ht]
  \begin{center}
    \includegraphics[width=.95\linewidth]{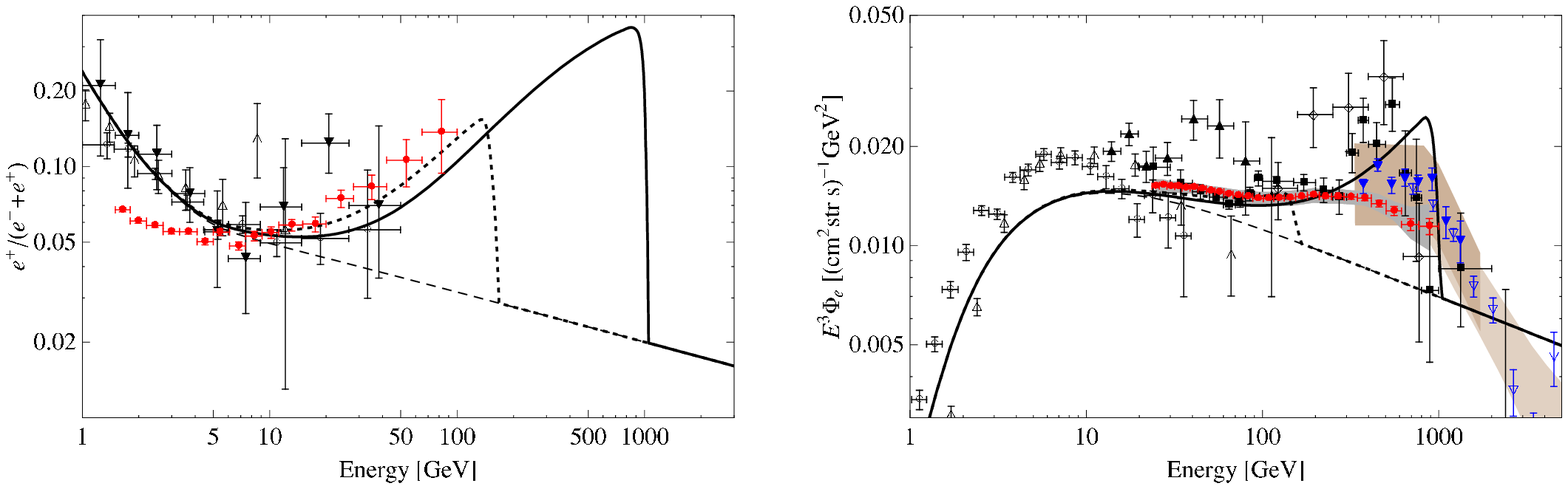}
    \includegraphics[width=.95\linewidth]{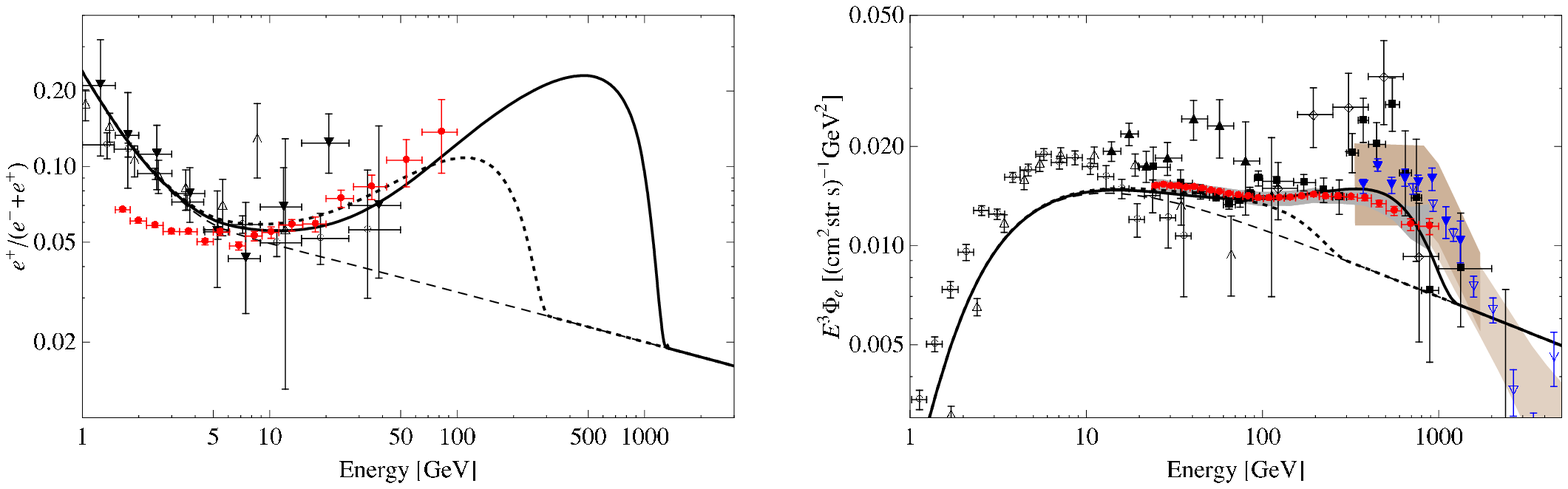}
    \includegraphics[width=.95\linewidth]{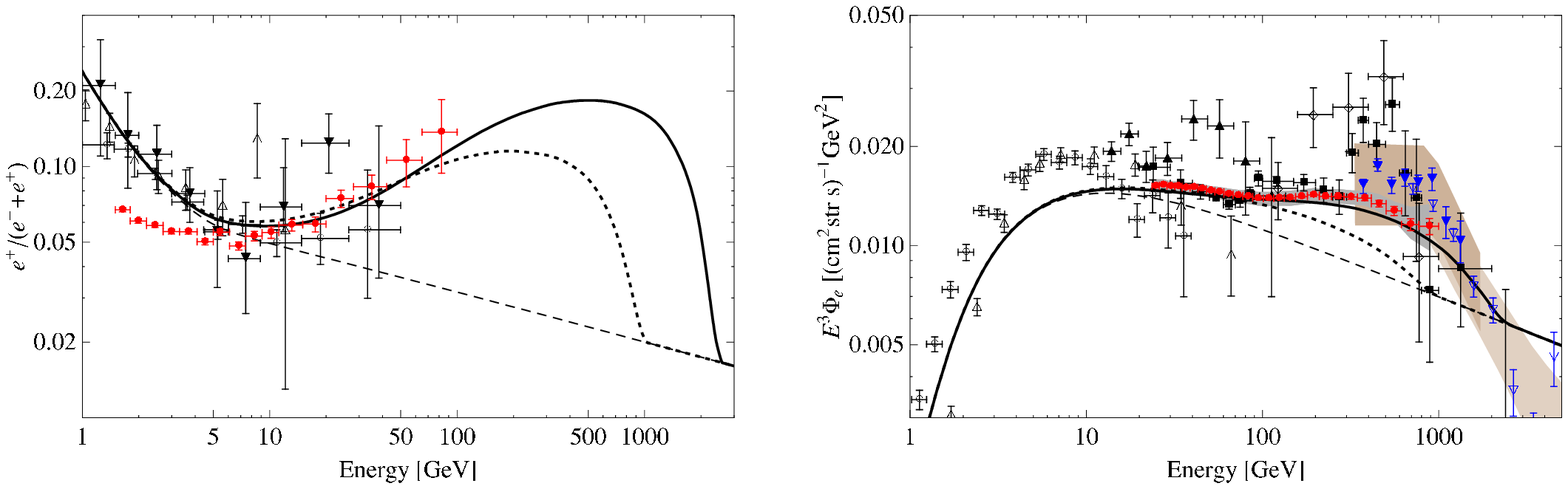}
    \vspace{-1cm}
  \end{center}
  \caption{Same as Fig.~\ref{fig:Znu}, but for the decay channels
  $\phi_\text{DM}\rightarrow \ell^+\ell^-$. {\it Upper panels:}
  $\phi_\text{DM}\rightarrow e^+e^-$ with $M_\text{DM}=2000\GeV$ (solid) and
  $300\GeV$ (dotted). {\it Middle panels:} $\phi_\text{DM}\rightarrow
  \mu^+\mu^-$ with $M_\text{DM}=2500\GeV$ (solid) and $600\GeV$ (dotted).
  {\it Lower panels:} $\phi_\text{DM}\rightarrow \tau^+\tau^-$ with
  $M_\text{DM}=5000\GeV$ (solid) and $2000\GeV$ (dotted).}
  \label{fig:3ll}
\end{figure}

\begin{figure}[ht]
  \begin{center}
    \includegraphics[width=.95\linewidth]{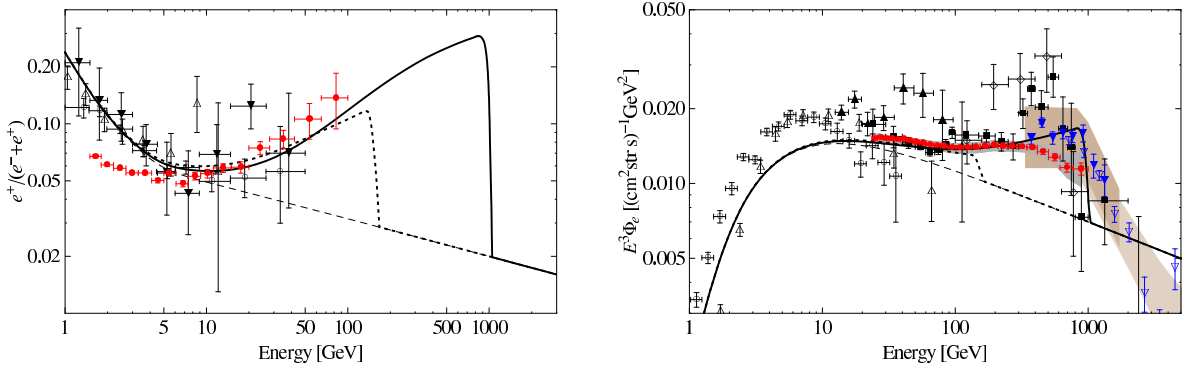}
    \vspace{-1cm}
  \end{center}
  \caption{Same as Fig.~\ref{fig:Znu}, but for the decay channel
  $\phi_\text{DM}\rightarrow \ell^+\ell^-$, democratic decay into three
  charged lepton flavors, with $M_\text{DM}=2000\GeV$ (solid) and $300\GeV$
  (dotted).}
  \label{fig:leplep}
\end{figure}

\begin{figure}[ht]
  \begin{center}
    \includegraphics[width=.45\linewidth]{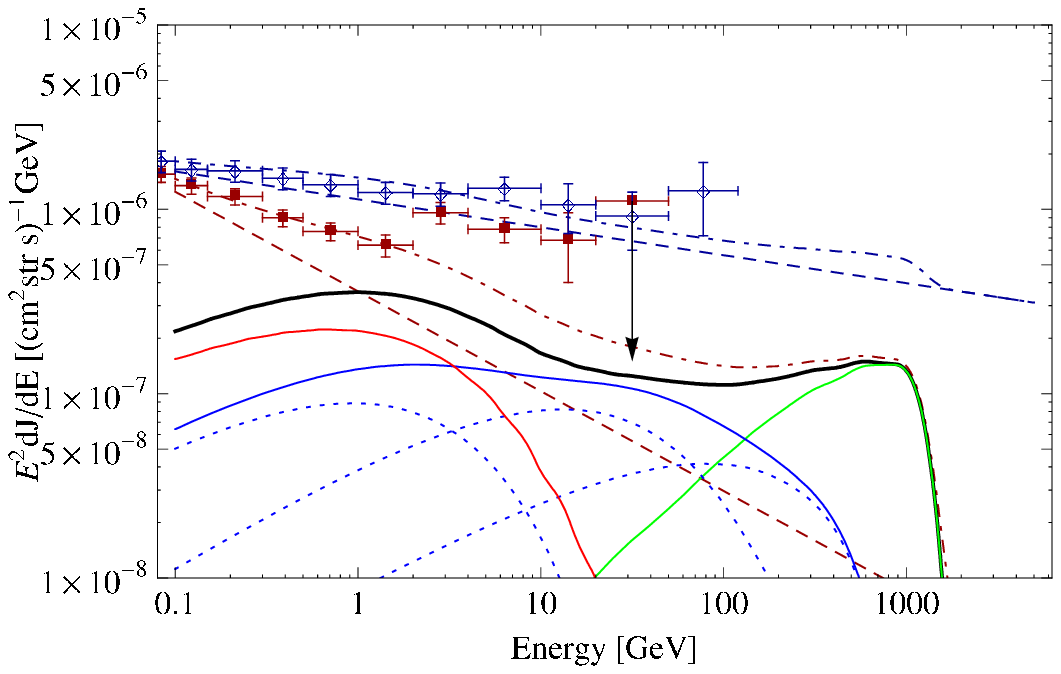}
    \includegraphics[width=.45\linewidth]{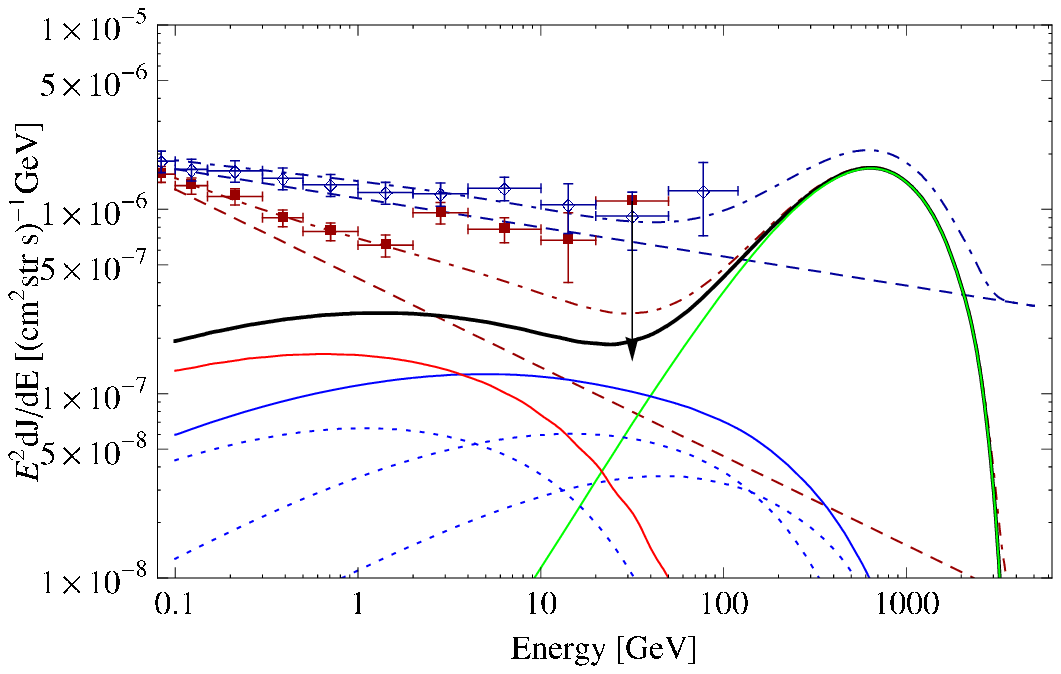}
    \vspace{-1cm}
  \end{center}
  \caption{Same as Fig.~\ref{fig:Wmu_addon}, but for
  $\phi_\text{DM}\rightarrow\mu^+\mu^-$ ({\it left panel}, with
  $M_\text{DM}=2500\GeV$) and $\phi_\text{DM}\rightarrow \tau^+\tau^-$ ({\it
  right panel}, with $M_\text{DM}=5000\GeV$).}
  \label{fig:ll_addon}
\end{figure}

Decays into harder electrons and positrons can arise in scenarios where the
scalar dark matter particle decays into a lepton-antilepton pair. We show in
Fig.~\ref{fig:3ll} the predictions for the positron fraction and the total
electron-plus-positron flux when the scalar dark matter particle decays into
fermions of the same generation, for dark matter masses between $M_{\rm
DM}=300\GeV$ and $5\TeV$. The decay $\phi_\text{DM}\rightarrow e^+ e^-$ can
explain the steep rise in the positron fraction observed by PAMELA. However,
it is apparent from Fig.~\ref{fig:3ll} that the dark matter decay into this
channel cannot be the origin of the Fermi excess in the total
electron-plus-positron flux.  The situation is similar if one considers
democratic decay into all three flavors as shown in Fig.~\ref{fig:leplep}.
Decays into softer electrons and positrons, as in the case when the dark
matter particles decay exclusively via $\phi_\text{DM}\rightarrow \mu^+
\mu^-$, are more promising.  In particular, a scalar dark matter particle with
a mass $M_\text{DM}\simeq 2500\GeV$ and a lifetime $\tau_\text{DM}\simeq
1.8\times10^{26}\s$, which decays exclusively into $\mu^+\mu^-$ pairs, can
reproduce both the steep rise in the spectrum observed by PAMELA and the total
electron-plus-positron spectrum measured by Fermi. The same holds true for
decay into tau flavors, with $M_\text{DM}\simeq 5000\GeV$ and
$\tau_\text{DM}\simeq0.9\times10^{26}\s$. For these two decay channels, we
also show the predictions for the gamma-ray fluxes in Fig.~\ref{fig:ll_addon},
which are again compatible with the present data and present a spectral shape
which could be visibly different from a power law, depending on the index of
the genuinely extragalactic contribution.

A summary of our results can by found in Tab.~\ref{tab:results}. Note that one
of the largest uncertainties that enter in the determination of the dark
matter lifetime comes from the determination of the local dark matter density
(see Ref.~\cite{Catena:2009mf} for a recent analysis) since this quantity is
inversely proportional to the corresponding flux of cosmic rays.

\begin{table}[ht]
  \centering
  \begin{tabular}{|c|c|c|}
    \hline
    Decay Channel & $M_\text{DM}$ [GeV] & $\tau_\text{DM}$
    [$10^{26}$s]\\\hline
    $\psi_\text{DM}\rightarrow \mu^+\mu^-\nu$ & 3500 & 1.1 \\
    $\psi_\text{DM}\rightarrow \ell^+\ell^-\nu$ & 2500 & 1.5\\
    $\phi_\text{DM}\rightarrow \mu^+\mu^-$ & 2500 & 1.8 \\
    $\phi_\text{DM}\rightarrow \tau^+\tau^-$ & 5000 & 0.9 \\
    $\psi_\text{DM}\rightarrow W^\pm\mu^\mp$ & 3000 & 2.1 \\\hline
  \end{tabular}
  \caption{Decay channels for fermionic and scalar dark matter,
  $\psi_\text{DM}$ and $\phi_\text{DM}$, respectively, that best fit the Fermi
  and PAMELA data for the MED propagation model and the NFW halo profile.  As
  discussed above, the dependence on the halo profile is negligible, while the
  dependence on the adopted propagation parameters is illustrated in Fig.
  \ref{fig:MINMEDMAX} for the decay $\psi_{\text{DM}} \rightarrow \mu^+ \mu^-
  \nu$.  The decay mode $\psi_{\text{DM}} \rightarrow W^\pm \mu^\mp$ is in
  tension with the PAMELA results on the antiproton-to-proton ratio, as
  mentioned in the text.}
  \label{tab:results}
\end{table}

\section{Conclusions} 
In some well-motivated dark matter scenarios, the dark matter particles are
unstable and decay with a lifetime much longer than the age of the Universe.
In this paper we have investigated whether the anomalies in the positron
fraction and the total electron-plus-positron flux reported by the PAMELA and
the Fermi LAT collaborations, respectively, could be interpreted as a
signature of the decay of dark matter particles. We have shown that some
decaying dark matter scenarios can indeed reproduce the energy spectra of the
positron fraction and the total flux reasonably well, while being at the same
time consistent with present measurements of the antiproton flux and the
diffuse extragalactic gamma-ray flux. The most promising decay channels for a
fermionic or a scalar dark matter particle are listed in
Tab.~\ref{tab:results}, where we also show the approximate mass and lifetime
which provide the best fit to the data. It should be borne in mind that the
astrophysical uncertainties in the propagation of cosmic rays and in the
determination of the background fluxes of electrons and positrons are still
large. Besides, the existence of a possibly large primary component of
electrons/positrons from astrophysical sources, such as pulsars, cannot be
precluded. Therefore, the precise values of the dark matter parameters can
vary. The present results can nevertheless be used as a guidance for building
models with decaying dark matter as an explanation of the PAMELA and Fermi
anomalies. 

Future measurements of the extragalactic diffuse gamma-ray flux by the Fermi
LAT will provide important information about the decaying dark matter
scenario. First, since the Earth is located far from the center of the Milky
Way halo, an anisotropy in the diffuse extragalactic gamma-ray flux is
expected which could be observed by the Fermi LAT~\cite{BBCI}. Moreover, all
scenarios in Tab.~\ref{tab:results} predict a departure from a simple power
law in the energy spectrum of the extragalactic diffuse background, the
deviation depending on the spectrum of the genuinely extragalactic
contribution originating presumably from AGN. The observation of such a
deviation would provide support for the decaying dark matter scenario and may
help to discriminate among the different possibilities in
Tab.~\ref{tab:results}.

\section*{Acknowledgements}

We are grateful to the anonymous referee for helpful comments. The work of AI
and DT was partially supported by the DFG cluster of excellence ``Origin and
Structure of the Universe.''

\bibliography{}
\end{document}